\documentclass{PoS}
\pdfoutput=1

\title{Optical Afterglows and IGM Attenuation}

\ShortTitle{IGM Attenuation: Photometric Corrections}

\author{\speaker{J.~Japelj}$^1$, A. Gomboc$^{12}$ and D. Kopa\v{c}$^1$\\
		\llap{$^1$}Faculty of Mathematics and Physics, University of Ljubljana, Jadranska ulica 19, SI-1000 Ljubljana, Slovenia\\
		\llap{$^2$}Centre of Excellence, SPACE-SI, A\v{s}ker\v{c}eva cesta 12, SI-1000 Ljubljana, Slovenia\\
		E-mail: \email{jure.japelj@fmf.uni-lj.si}, \email{andreja.gomboc@fmf.uni-lj.si}, \email{drejc.kopac@fmf.uni-lj.si}}

\abstract{Because of their cosmological origin, gamma-ray burst (GRB) optical afterglows are attenuated when they pass intergalactic absorbers in the GRB line-of-sight. Without the knowledge of the number of absorbers and their physical properties, the effect of absorption on the observed magnitudes can not be determined precisely. Different methods have been applied in order to correct for this effect statistically, either using semi-analytical calculations or numerical simulations. We follow these works and present the expected magnitude corrections as a function of redshift for a set of filters most commonly used in the scientific community. The results are publically available on the web (\textit{http://igmac.fmf.uni-lj.si}).}

\FullConference{Gamma-Ray Bursts 2012 Conference -GRB2012,\\
		May 07-11, 2012\\
		Munich, Germany}

\begin{document}

\section{Introduction}
Spectra of distant emitters (galaxies, quasars, gamma-ray burst afterglows) reveal a rich set of absorption features - a result of light passing through a number of intergalactic gaseous clouds (intergalactic medium - IGM). While various absorption features can be found in the observed spectra (e.g., MgII, CIV, OVI, HeII and others), by far the most abundant absorber is neutral hydrogen. Depending on the physical properties of clouds (their redshift, hydrogen column density and temperature), the absorption features take various forms (e.g., Lyman forest, Lyman limit systems, damped Lyman systems). For a review on IGM absorbers, see \cite{mo}. 

Neglecting absorption from all other elements except hydrogen, the light from a distant emitter is partially absorbed blueward of the redshifted Ly$\alpha$ line. Any photometric observation in these wavelengths results in an uncertainty: how much light has been lost due to the IGM attenuation?

Without simultaneous spectroscopic and photometric observations the photometric correction can only be applied statistically, either with the help of semi-analytical tools or simulations. Even though the attenuation values obtained in that way are based on average properties of IGM and may differ substantially from actual losses of light in some cases (e.g., if a damped Lyman system is in the line-of-sight, the attenuation is going to be underestimated), the results are still useful and informative.

\section{Methods}
In order to determine the IGM effects on light emitted from a distant source, two methods are employed.

The most widely used method is presented in \cite{madau1995}, where measured properties of the absorbers (i.e., number evolution with redshift, column density distribution, temperature of the clouds) were used to construct a semi-analytical model of the transmission function averaged over many different lines-of sight - $\Phi(\lambda)$. When the transmisson function is known, the attenuation of light observed through a filter with known filter transmission function $T(\lambda)$ can be computed as:
\begin{equation}
\Delta \textrm{mag} = -1.086\ln Q,
\end{equation}
where $Q$ is the mean transmission function, averaged over a filter bandpass:
\begin{equation}
Q = \frac{\int \Phi(\lambda)T(\lambda)d\lambda}{\int \Phi(\lambda)d\lambda}.
\end{equation}
The approach relies on the assumption that the observed flux from a distant emitter at redshift $z_{\textrm{em}}$ is a product of the emitted flux and the mean transmission function.

The other approach mimics the measurement process. The photometric attenuation is first calculated for each line-of-sight separately and then the statistical analysis is applied to the obtained results. This can be accomplished with a Monte Carlo simulation and has been applied in many works (\cite{moller}, \cite{bershady}, \cite{garcia}, \cite{inoue}). Different works mainly differ in the choice of parametrisation describing IGM properties. Absorption properties in a particular line-of-sight are determined by a set of parameters: 
\begin{equation}
\lbrace z_{i}, N_{\mathrm{HI,i}}, b_{\mathrm{i}}\rbrace_{i=1,...,N_{\mathrm{abs}}},
\end{equation}
where $N_{\mathrm{abs}}$ is the number of absorbing clouds in the line-of-sight, each described by a unique set of parameters: redshift $z_{\mathrm{i}}$, neutral hydrogen column density $N_{\mathrm{HI,i}}$ and Doppler parameter $b_{\mathrm{i}}$. The idea is to simulate a number of lines-of-sight (up to redshift $z_{\mathrm{em}}$), where $N_{\mathrm{abs}}$, $z_{\mathrm{i}}$, $N_{\mathrm{HI,i}}$ and $b_{\mathrm{i}}$ are generated according to empirical distributions. Magnitude increment (corresponding to a given filter transmission curve) is calculated for each line-of-sight. The resulting distribution of increments is then statistically analysed (see below).

We perform the Monte Carlo simulation by using two different models describing IGM properties: one is the same as the one used in \cite{madau1995} (model A), while the other is taken from Table 1 in \cite{inoue} (model B). The latter model is more detailed than the former: ($i$) it includes two observed breaks in the absorbers' number evolution with redshift (e.g., \cite{weymann} \cite{fan}), ($ii$) it includes a break in the column density distribution (i.e., larger fraction of strong absorbers) \cite{inoue} and ($iii$) the Doppler parameter is assumed to be distributed according to eq. 3 in \cite{kim}.

\begin{figure}[!b]
\begin{center}$
\begin{array}{cc}
\includegraphics[width=7.5cm]{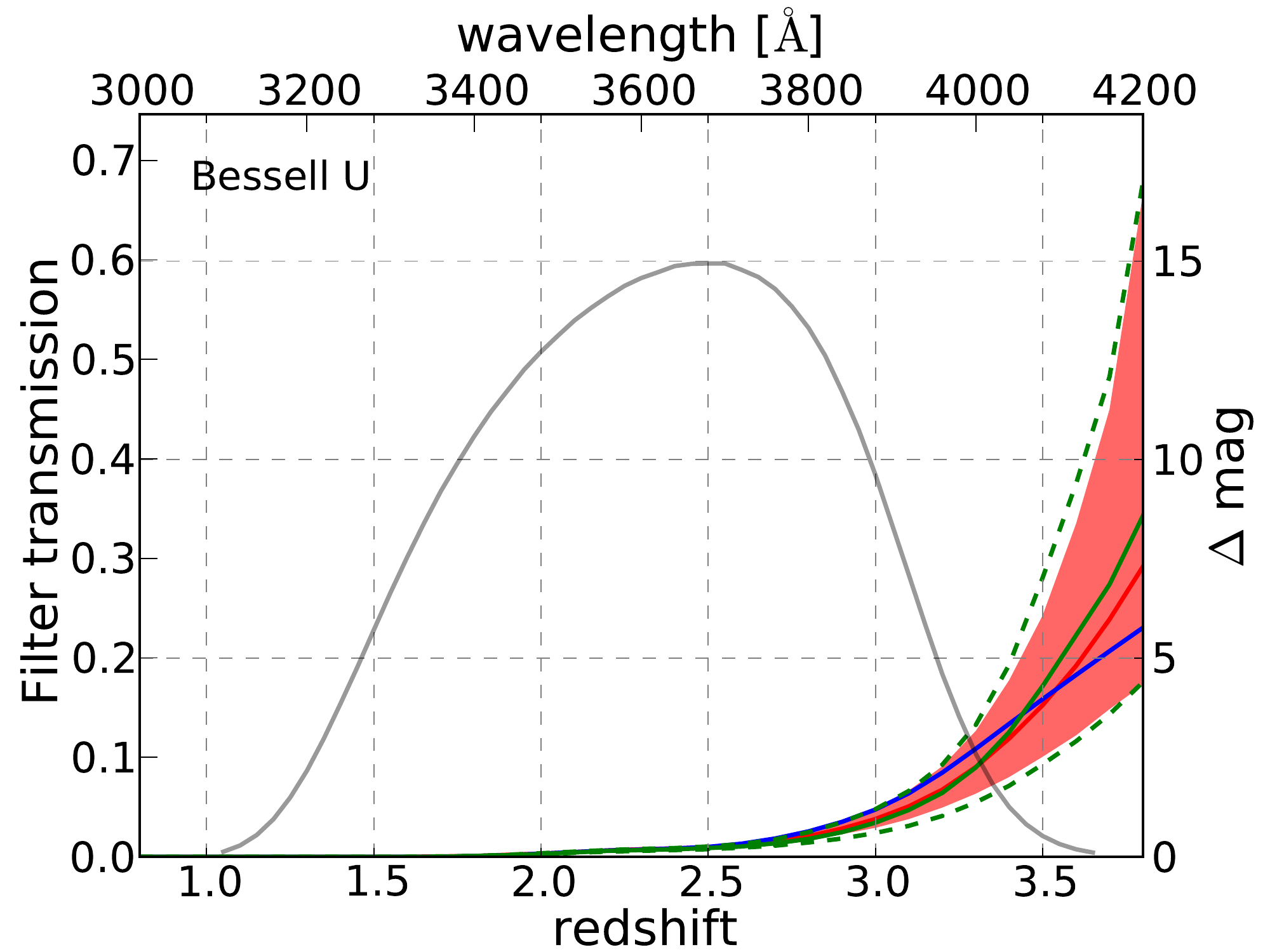}&
\includegraphics[width=7.5cm]{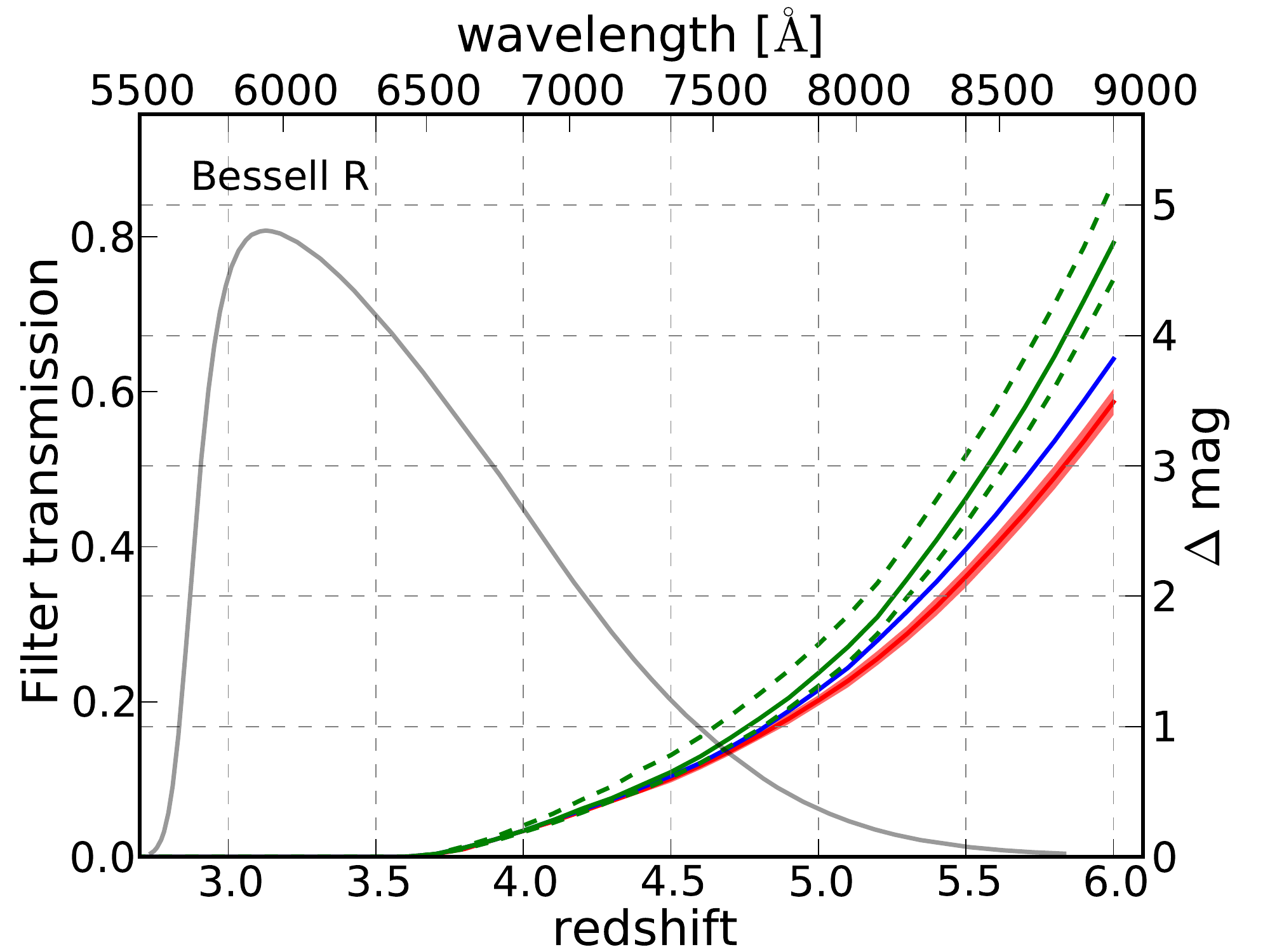}
\end{array}$
\end{center}
\caption{Magnitude correction for Bessell $U$ and $R$ filters as a function of emitter's redshift: semi-analytical model (blue), model A (red; shaded area corresponds to the area between $\pm$ 1 $\sigma$ values), model B (green; dashed lines correspond to $\pm$ 1 $\sigma$ values). Filter transmission curve is given for reference (grey line). The redshift and wavelength scales are completely independent!}
\label{fig1}
\end{figure}

\section{Results}
We compute the photometric corrections for the redshift interval $z_{\mathrm{em}}$ = 0.5 - 6.0 with a resolution of $dz_{\mathrm{em}}$ = 0.1 for 10000 lines-of-sight. IGM absorbers are known to have a wide range of column density values (from $10^{12}-10^{22} \mathrm{cm}^{-2}$) \cite{mo}. Attenuation for different lines-of-sight therefore differs a lot. The distribution of magnitude increments for an emitter at specific redshift is not normal in general (as it turns out, taking the logarithm of the magnitudes does not result in a log-normal distribution). Therefore we compute a 50$\%$ ($\pm$ 34.13$\%$) quantile (corresponding to median $\pm$ 1 $\sigma$) for each distribution. 

Computed photometric corrections for the case of Bessell $U$  and Bessell $R$ filters are presented in Figure 1. The graphs show computed median magnitude correction as a function of redshift for: semi-analytical model (blue), simulation results using model A (red) and simulation results using model B (green). It is interesting to note that the results obtained by using semi-analytical model and model B differ, even though the assumed parametrisation is the same - the difference has been also observed in previous works \cite{bershady}\cite{garcia}. From the two figures one can notice a large uncertainty in the case of Bessell $U$ filter at high redshifts compared to the uncertainty in corrections for Bessell $R$ filter. Closer examination reveals that the uncertainty starts rising when most of the filter transmission resides below the redshifted Lyman edge (i.e., $(1 + z_{\mathrm{em}})\times 91.3$ nm), where the strength of continuous opacity for high energy photons is highly dependent on the $N_{\mathrm{HI}}$ of the absorber (e.g., \cite{mo}\cite{madau1995}\cite{garcia}) - hence the large differences among different lines-of-sight. In addition to wavelength coverage, amplitude of the uncertainty also depends on the width and shape of the filter transmission function.
%Due to large uncertainties at high redshifts, the corrections for Bessell $U$ filter have been computed only up to the redshift for which the average transmission curve in the wavelength range, covered by the filter, is approximately zero.

\section{Summary}

We compute photometric corrections for a set of filters most commonly used in the scientific community. The results are publically available on the IGM Attenuation Correction web page$\footnote{http://igmac.fmf.uni-lj.si}$ and are intended to serve as a quick check on the expected average attenuation of light for distant objects at known redshifts.

\end{document}